# 10 Inventions on Command Buttons in a Graphical User Interface

**Umakant Mishra**

Bangalore, India

http://umakantm.blogspot.in

**Contents**

## 1. Introduction

The recent trend in software development shows more and more importance on graphical user interface. A graphical user interface is often seen as just indispensable. There are several important components of a graphical user interface, such as, menu, listbox, toolbox, command button, scrollbar, tree navigator, dialog box etc. A command button is a very fundamental element of a graphical user interface. A command button provides an easy access to a command or function in the system. The user can click on the command button with a pointing device to execute the operation.

A command button may contain a textual label or a graphic image or both. It may be static or animated. There can be many different features to make a command button attractive and effective. As command button is a typical GUI element, most improvement on GUI in general will also be applicable to command buttons. Besides, there are also inventions to improve various aspects of command buttons in specific. This article illustrates 10 selected inventions from US patent database. Each invention is followed by a TRIZ based analysis in brief.

## 2. Inventions on Command buttons

### 2.1 Method for adding a graphical user interface to a command line application (5617527)

**Background problem**

Many programs use a command line prompt where the user can submit a command to be processed. This has the drawback that the user has to remember the commands; besides there is a possibility of spelling errors while typing the commands. Moreover, the users also find difficulties with two different "standards", viz., the GUI and command line. There is a need to convert the pre-existing programs for user convenience and to protect the investment base.

**Solution provided by the invention**

Patent 5617527 (invented by Kressin et al., assigned to IBM, issued April 1997) disclosed a method of adding graphical user interface to existing command line applications. According to invention a set of GUI "buttons" are set up for different command line commands. The user can operate the buttons with mouse and run the associated command line command.

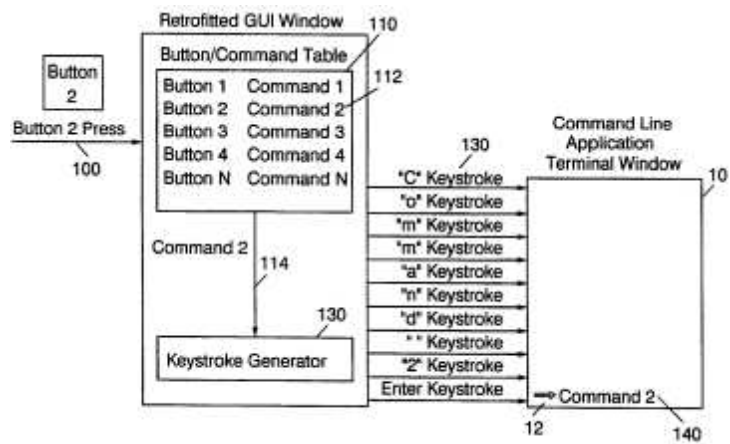

This method avoids typing of the commands from command line interface by converting them into GUI buttons. Thus the invention displays an array of "buttons" each of which is associated with different commands. The user selects a button with mouse, which sends the command to the application to execute.

**TRIZ based analysis**

The invention converts the command line commands and parameters into GUI buttons. The buttons are operated through mouse buttons avoiding typing of commands (Principle-28: Mechanics substitution).

**2.2 Closely integrated key icons in a graphical interface (5694562)**

**Background problem**

Generally the keyboards have a set of function keys that are associated with different functions defined in specific applications. With the popularity of GUI users are habituated with operating through a mouse and don't prefer to remove hand from the mouse and press a function key. It is desirable to press a function key by without lifting the hand from the mouse.

**Solution provided by the invention**

Patent 5694562 (invented by Thomas Fisher, issued Dec 1997) discloses a graphical user interface that contains buttons resembling with the function keys on the keyboard. The user can invoke a function either by pressing a function key or by clicking the corresponding function key button on the screen. In either case, the icon becomes animated to acknowledge invocation of the function.

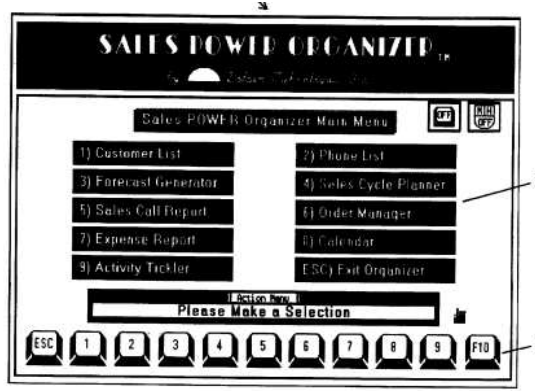

**TRIZ based analysis**

The invention invokes a function key function by using a mouse and without pressing the function key on the keyboard. (Principle-28: Mechanics substitution).

The invention uses the picture of function keys in the form of buttons that provide the same functionality as the function keys (Principle-26: Copying).

## 2.3 GUI pushbutton with multi-function mini-button (5736985)

**Background problem**

With the popularity of graphical user interface more and more functions are integrated into the graphical user interface. But when more are more buttons are added to the screen, the interface becomes crowded, confusing and less usable. This problem is addressed to some extent by a multi-function push button. The multi-function push button contains a smaller mini-button inside a larger button. If the user clicks anywhere on the larger pushbutton's area, one function is invoked, but if the user clicks on the mini-button, then both the functions are invoked, i.e., first the large button function and then the mini-button function.

Although this mechanism of a multi-function push button is a substantial improvement on a conventional push button, it is not intuitively obvious which part of the pushbutton would perform both functions and which part would perform only one.

**Solution provided by the invention**

Patent 5736985 (invented by Lection et al., assigned by IBM, issued Apr 1998) provided a multifunction pushbutton, where the secondary function can be one of many alternatives.

According to the invention if the pointer is not located within the mini-button then the application invokes the primary functions and terminates the secondary function. If the pointer is located within the mini-button then the application first invokes the primary function for the pushbutton and then invokes the secondary function for the mini-button. If there are multiple mini-button functions available then the function represented by the icon is incremented to represent the next mini-button function available sequentially.

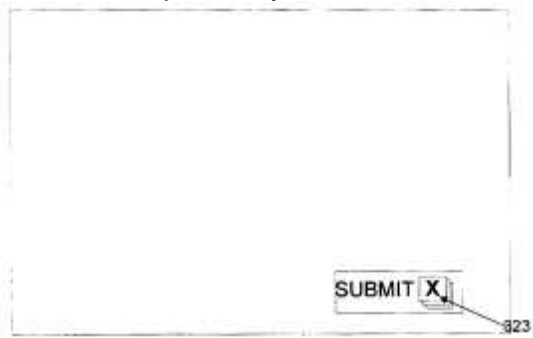

This method allows more number of functions in less number of buttons so that the screen is not cluttered.

**TRIZ based analysis**

The invention puts a mini-button inside a push button to save space and to achieve even two functions in a single click (Principle-7: Nested doll).

The invention allows multiple functions to be assigned to the same mini-button (Principle-6: Universality, Principle-7: Nested Doll).

## 2.4 Graphical user interface with icons having expandable descriptors (5748927)

**Background**

There are problems associated with displaying large number of icons on the user interface. One is that the icons don't have enough space to be displayed. Secondly, the user also fails to remember the function associated with each particular icon. There is a need to solve this problem.

**Solution provided by the invention**

Patent 5748927 (Invented by Stein et al., Assigned to Apple Computers, May 98) discloses a method of displaying expanded description for the icons. According to the invention, the icons consume as little space as possible by arranging them proximate to one another. When the cursor is not placed over the icon, only a part of the icon descriptor text is displayed within the width of the icon. But when the cursor is placed over an icon, the icon's descriptor text is accentuated to provide a more complete description.

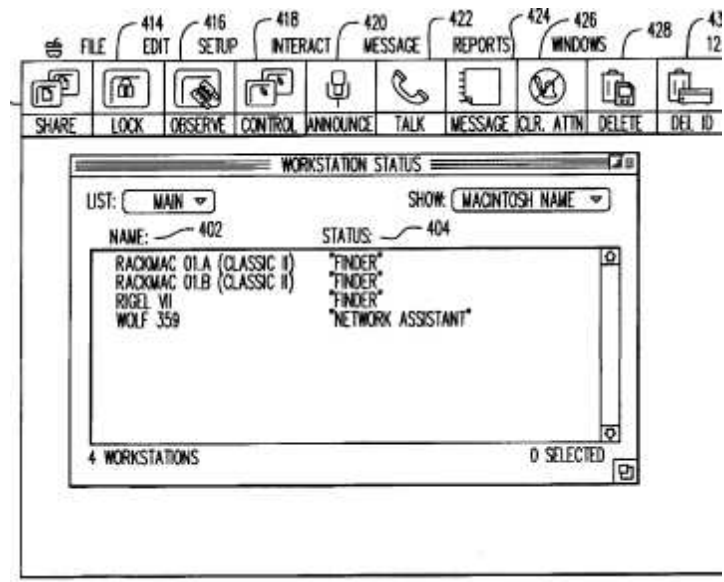

**TRIZ based analysis**

The invention displays a small description or partial description of the button to that fits within the width of the button (Principle-16: Partial or excessive action).

When the cursor is placed over the icon the descriptor text is accentuated to display a full description (Principle- Dynamize).

## 2.5 Menu control in a graphical user interface (5828376)

**Background problem**

In a graphical user interface, the commands are normally executed from a dropdown menu or from a toolbar. However both of them have their own shortcomings. The menu needs complex navigation before reaching the desired option and the toolbar occupies the real estate of the screen. There is a need to improve the method of accessing and executing commands by overcoming the prior art methods.

**Solution provided by the invention**

Patent 5828376 (Invented by Solimene et al., Assigned to J.D. Edwards World Source Company, Oct 98) provides a "hyperbutton", a "context sensitive pop-up menu" and a "menu control editor" for easy access to menu options in a GUI.

The hyperbutton displays a user selected default hyperitem from the hyperitems of the menubar. The user can change the default hyperitem using a pop-up menu displayed by using (say) right mouse button. A hyperbutton editor allows the user to configure the hyperbutton and context sensitive pop-up menu be defining the associated hyperitems and their attributes.

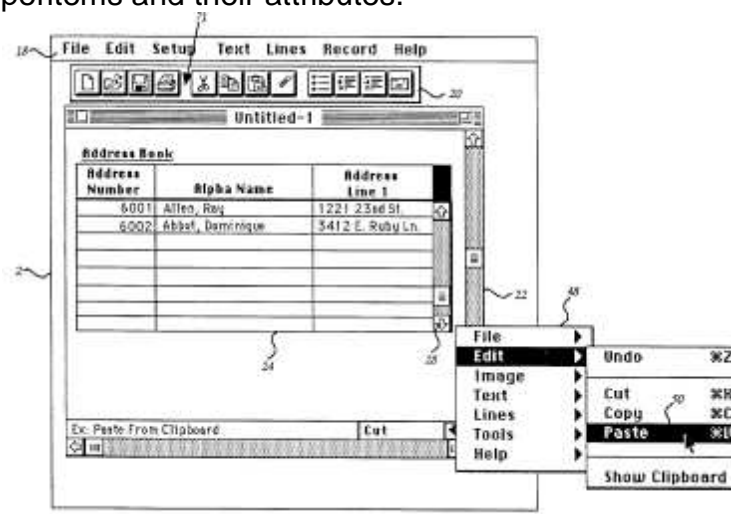

The hyperbutton is supposed to be used in addition to the standard menubar and toolbar features in a GUI.

**TRIZ based analysis**

The hyperbutton displays the default hyperitem from the hyperitems of the menubar that helps user to directly execute from hyperbutton instead of navigating the options through the menu (Principle-10: Prior Action).

The user can change the default hyperitem (Principle-15: Dynamize).

## 2.6 Method and system for presenting a plurality of animated display objects to a user for selection on a graphical user interface in a data processing system (5838316)

### Background problem
The graphical user interface makes a system user friendly. A user can easily interact with the computer without knowing any complex commands. It is necessary to make the GUI more interesting that will encourage the user to explore more into the system and should be easy to remember too.

### Solution provided by the invention
Patent 5838316 (invented by Arruza, assigned by IBM, Nov 1998) discloses a method of presenting animated display objects to a user for selection. According to the invention a plurality of animated display objects are simultaneously displayed by a graphical user interface allowing the user to view all and select any one of them.

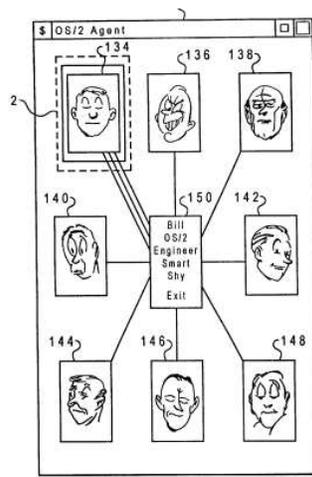

### TRIZ based analysis
The invention improves the conventional GUI to make more attractive and the items easy to remember (Principle-38: Enrich).

The invention presents animated display objects instead of conventional static display objects (Principle-15: Dynamize).

## 2.7 Graphical user interfaces having animated control elements (5880729)

### Background problem
In a graphical user interface, the control elements are typically displayed in a static state. Some buttons has two display states, i.e., normal and depressed. With the advancement of GUI, it is desirable to provide advanced visual characteristics to the buttons.

**Solution provided by the invention**

Patent 5880729 (invented by Johnston Jr., et al., assigned by Apple Computer Inc., issued Mar 1999) provides an enhanced visual appearance to the graphical user interface. The control elements in the graphical user interface are associated with at least two states; a first display state is associated with a first functional state and the second display state is associated with a second functional state. When transitioning between the states, an animated visual effect is provided to the control element.

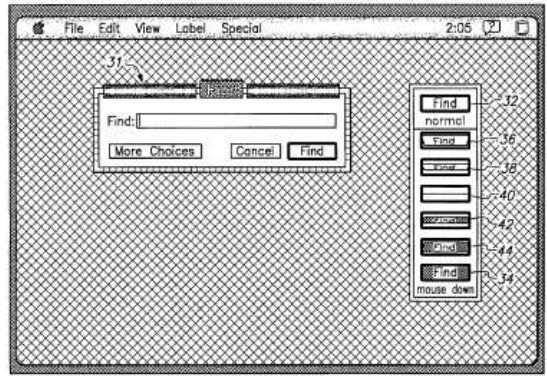

**TRIZ based analysis**

The invention enhances the visual effect of buttons (and similar control elements) by animating them when selected (Principle-38: Enrich, Principle-15: Dynamize).

**2.8 Method, apparatus and computer program product for graphical user interface control and generating a multitool icon (6091416)**

**Background problem**

Many software products include multi-function toolbox that contains multiple tools like a line drawing tool, a paintbrush and an eraser. The user typically selects a tool with a mouse and use the tool on the worksheet. The user then selects another tool and thus moves back and forth between the toolbox and the worksheet.

**Solution provided by the invention**

Patent 6091416 (invented by Cragun, assigned by IBM, Jul 2000) provides an improved method that efficiently and effectively facilitates user control to selectively access multiple user-selected tools. According to the invention when the user selects a tool, a multitool icon is built and displayed responsive to the user-selected tool. The multitool icon is displayed within a selected display screen area.

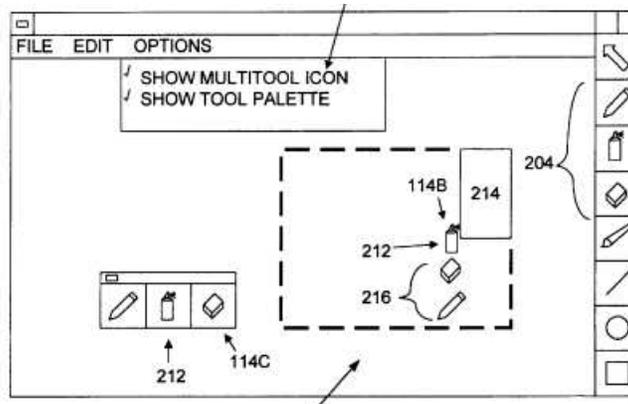

The user selects tools from the multitool icon and toggles back and forth within the multitool icon without moving the mouse pointer away from the work area.

**TRIZ based analysis**

The multitool icon contains only the user-selected tools and does not contain all other tools in the toolbox (Principle-2: Taking out). This facilitates quick selection.

Instead of user moving the cursor to the toolbox the multitool icon is displayed near the cursor. (Principle-13: Other way round).

**2.9 Method for displaying controls in a system using a graphical user interface (6384849)**

**Background problem**

The GUI of an application program provides controls and functions through menus and toolbars. The user can choose the functions either through the menu or from the toolbar. Although toolbar and dropdown menus both provide ways to display controls, they look and feel very different. While toolbars have rich interactive controls, the drop down menu contains simple text strings. Both of them are treated differently and may contain different functions.

There is a need of an improved command bar that allows all controls to be included in either menu type containers or toolbar type containers.

**Solution provided by the invention**

Marcos et al. invented a method of displaying controls through a command bar (Patent 6384849, Assigned to Microsoft, May 02). According to the invention, the items may be displayed as both menu-like containers and toolbar like containers. The details of the control items are stored in a database. The data items include the Identification and description of the command bar along with their display state, i.e., whether menu-like or toolbar-like. The items are enabled for menu and/or for toolbar depending on their frequency of use.

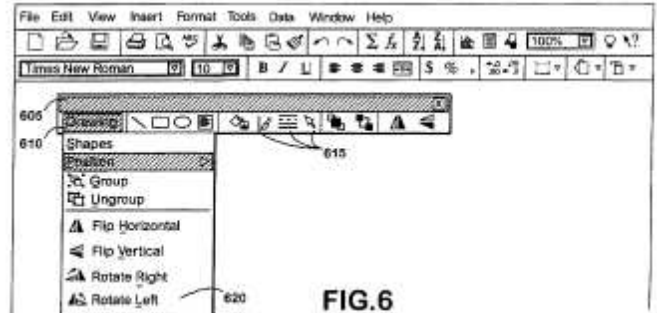

FIG.6

This invention of command bars integrate the features of both menubars and toolbars and provides methods for customizing them.

**TRIZ based analysis**

The command bar contains the functions and features of both menubar and toolbar (Principle-40: Composite).

The invention allows adding and removing menu popup from the command bar (Principle-15: Dynamize).

**2.10 Graphical user interface for selection of options within mutually exclusive subsets (6535229)**

**Background problem**

The items in a checkbox are not mutually exclusive. A user can selectively check or uncheck a single or multiple options in a checkbox. Compared to that the options in a radio button are mutual exclusive. The user can select only one option in a radio box. When one option is selected the other options are automatically deselected.

Although the check box lists and radio button lists are appropriate in many situations, there are certain situations where these controls are awkward and inconvenient. There is a need for a better virtual control.

**Solution provided by the invention**

Patent 6535229 (invented by Kraft, assigned to IBM, issued March 2003) provides a virtual control for inputting data to the computer. The control presents a set of selectable options to a user and allows the user to select items from the set of options. The set of selectable items is partitioned into subsets. Items within the subset may be selected or de-selected independent of each other, while items in different subsets may have interdependent settings. The selection of any item in one subset forces all items in all other subsets to be de-selected, so that the selection of items in distinct subsets are mutually exclusive.

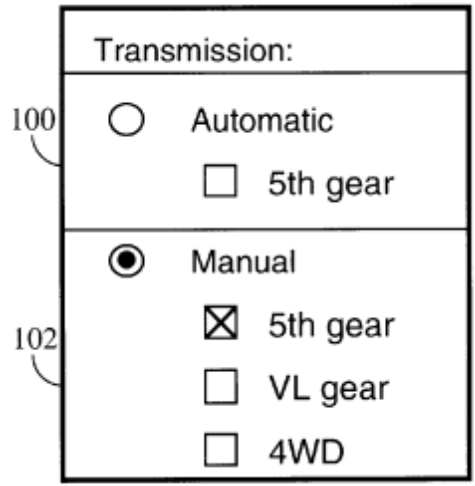

This GUI control ensures self-consistency of information in cases where the conventional radio button and check box GUI controls are inadequate.

**TRIZ based analysis**

The invention combines the functionality of both a radio box control and check box control into one single control with added features like self-consistency (Principle-40: Composite).

The items in the list are grouped in sub-sets. (Princple-1: Segmentation).

The invented control maintains self-consistency by unselecting all other subsets when any option in one subset is selected (Principle-25: Self service).

## 3. Summary and Conclusion

A basic command button can be improved to provide better visual effects, improved selection mechanism and added features of advantage. Each of the above ten inventions analyzed above try to improve the feature or function of a command button by some way or other. We have used TRIZ principles to simplify the solution process. The principles can be applied to find further solutions for future improvements.

## Reference:

1. US Patent 5617527, "Method for adding a graphical user interface to a command line application", invented by Kressin et al., assigned to IBM, issued April 1997.

2. US Patent 5694562, "Closely integrated key icons in a graphical interface", invented by Thomas Fisher, issued Dec 1997.

3. US Patent 5736985, "GUI pushbutton with multi-function mini-button", invented by Lection et al., assigned by IBM, issued Apr 1998.

4. US Patent 5748927, "Graphical user interface with icons having expandable descriptors", Invented by Stein et al., Assigned to Apple Computers, May 98.

5. US Patent 5828376, "Menu control in a graphical user interface", Invented by Solimene et al., Assigned to J.D. Edwards World Source Company, Oct 98.

6. US Patent 5838316, "Method and system for presenting a plurality of animated display objects to a user for selection on a graphical user interface in a data processing system", invented by Arruza, assigned by IBM, Nov 1998.

7. US Patent 5880729, "Graphical user interfaces having animated control elements", invented by Johnston Jr., et al., assigned by Apple Computer Inc., issued Mar 1999.

8. US Patent 6091416, "Method, apparatus and computer program product for graphical user interface control and generating a multitool icon", invented by Cragun, assigned by IBM, Jul 2000.

9. US Patent 6384849, "Method for displaying controls in a system using a graphical user interface", invented by Marcos et al., Assigned to Microsoft, May 02

10. US Patent 6535229, "Graphical user interface for selection of options within mutually exclusive subsets", (invented by Kraft, assigned to IBM, issued March 2003.

11. US Patent and Trademark Office (USPTO) site, http://www.uspto.gov/